\documentclass{PoS}
\usepackage{verbatim}
\usepackage[hang,small,bf]{caption}
\title{Development and Performance of spark-resistant Micromegas Detectors}

\ShortTitle{Development and Performance of spark-resistant Micromegas Detectors}

\author{\speaker{George Iakovidis}\\
        National Technical Univ. of Athens, Greece\\
       E-mail: \email{George.Iakovidis@cern.ch}}
\author{{Konstantinos Karakostas}\\
	National Technical Univ. of Athens, Greece\\
       E-mail: \email{Konstantinos.Karakostas@cern.ch}}
\author{{Matthias Schott}\\
	CERN, Switzerland\\
       E-mail: \email{Matthias.Schott@cern.ch}}

\abstract{The Muon ATLAS MicroMegas Activity (MAMMA) focuses on the development and testing of large-area muon detectors based on the bulk-Micromegas technology. These detectors are candidates for the upgrade of the ATLAS Muon System in view of the luminosity upgrade of Large Hadron Collider at CERN (sLHC). They will combine trigger and precision measurement capability in a single device. A novel protection scheme using resistive strips above the readout electrode has been developed. The response and sparking properties of resistive Micromegas detectors were successfully tested in a mixed (neutron and gamma) high radiation field, in a X-ray test facility, in hadron beams, and in the ATLAS cavern. Finally, we introduced a 2-dimensional readout structure in the resistive Micromegas and studied the detector response with X-rays.}

\FullConference{XXIst International Europhysics Conference on High Energy Physics\\
		 21-27 July 2011\\
		 Grenoble, Rhones Alpes France}

\begin{document}

\section{Introduction}

The Micromegas (Micro-MEsh Gaseous Structure) detectors have been invented for the detection of ionizing particles in experimental physics, in particular in particle and nuclear physics. It was first proposed in 1996 \cite{Mic1}; its basic operation principle is illustrated in Fig. 1. A planar drift electrode is placed few mm above a readout electrode. The gap is filled with ionization gas. In addition, a metal mesh is placed $\sim0.1$ mm above the readout electrode. The region between drift electrode and mesh is called the drift region, while the region between mesh and readout electrodes is called the amplification region. Both the mesh and the drift electrode are set at negative high voltage, so that a electric field of $\sim600$ V/cm is present in the drift region and a field of $\sim50$ kV/cm is present in the amplification region. The readout electrodes are set to ground potential.
Charged particles transversing the drift region ionize the gas. The resulting ionization electrons drift towards the mesh with a drift velocity of 5 cm/$\mu$s. The mesh itself appears transparent to the ionization electrons when the electric field in the amplification region is much larger than that in the drift region. Once reaching the amplification region, the ionization electrons cause a cascade of secondary electrons (avalanche) leading to a large amplification factor, which can be measured by the readout electrodes. A significant step in the development of Micromegas detectors was achieved in 2006 and its known as bulk-Micromegas technology. A detailed description can be found in \cite{Mic2}.

\section{Resistive Chambers}

The thin amplification region together with its high electric field implies a large risk of sparking. Sparks can cause damage to the detector itself, on the underlying electronics, but lead also to significant dead-times. This serious disadvantage was overcome recently, with the development of spark resistant Micromegas chambers by the MAMMA group \cite{Mic3}.  The resistive Micromegas developed by MAMMA group has separate resistive strips rather than a continuous resistive layer to avoid charge spreading across several readout strips and to keep the area affected by a discharge as small as possible. The resistive strips are separated by an insulating layer from the readout strips and individually grounded through a large resistance. The Micromegas structure is built on top of the resistive strips. It employs a woven stainless steel mesh which is kept at a distance of 128 $\mu$m from the resistive strips by means of small pillars (Fig. 1). Above the amplification mesh, at a distance of 4 or 5 mm, another stainless steel mesh  serves as drift electrode. The signal on the readout strips is then capacitively coupled to resistive strips. It has been shown that this design provides a spark-resistant layout for Micromegas chambers even in very high flux environments \cite{Mic4}.

\begin{figure} [!hp]
	\begin{center}
		\includegraphics[scale=0.5]{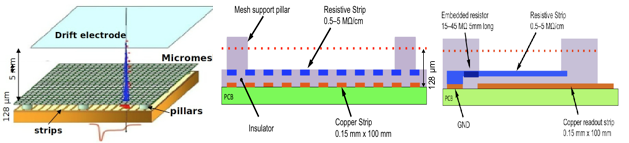}
				\setlength{\captionmargin}{10pt}
		\caption[Small]{Resistive Micrommegas Layout.}
	\end{center}
\end{figure}

The basic Micromegas design can be easily extended to a two-dimensional readout. The readout strips in the x-direction are placed parallel to the resistive strips, while the readout-strips in the y-direction are placed perpendicular. All strips are separated by isolation material. The signal on the readout strips is again capacitively coupled to resistive strips. Hence it is expected that the induced signal on the x-strips is smaller then the signal on the y-readout strips due to the larger distance to the resistive strips and screening effects. In order to ensure that the induced charge in both layers is of similar magnitude the lower readout-strips should be wider.

We present here preliminary results on the performance of spark resistant Micromegas chambers in a beam of neutrons with a flux of $10^6 Hz/cm^{2}$. The detectors have been operated with three Ar:CO$_2$ gas mixtures, with 80:20, 85:15 and 93:7. Fig. 2 shows a comparison of the the high voltage drop in case of sparks and the current that chamber draws for the bulk Micromegas on the left and resistive one on the right.  Only a few sparks per second were observed in a beam with 1.5$\cdot$10$^6$ neutrons$/$cm$^2/$s. Hence, the spark signal is reduced by a factor of 1000 compared to a standard Micromegas. The spark rate was found four times higher with the 80:20 compared to the 93:7 Ar:CO$_2$ gas mixture. The neutron interaction rate was found independent of the gas.

\begin{figure} [!htp]
	\begin{center}
		\includegraphics[scale=0.4]{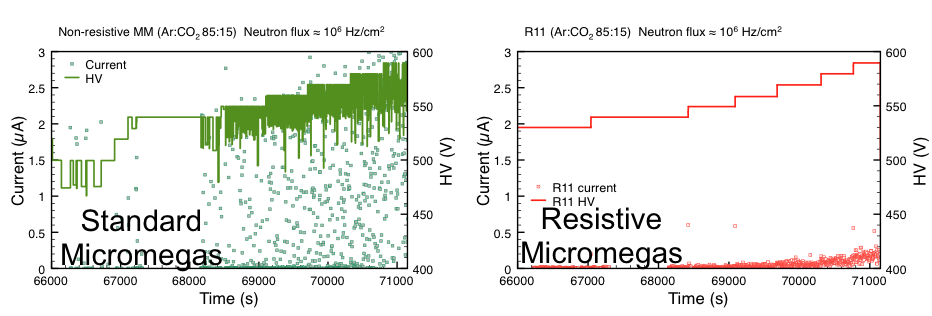}
				\setlength{\captionmargin}{10pt}
		\caption[Small]{Performance of standard (left) and resistive (right) Micromegas chambers.}
	\end{center}
\end{figure}


\begin{thebibliography}{99}
\bibitem{Mic1} I. Giomataris et al.,: Micro-Pattern Gaseous Detectors, Nucl. Instrum. Methods A 376 (1996) 29
\bibitem{Mic2} I. Giomataris et al., Micromegas in a bulk; Nucl Instrum. Methods, A560 2006, PP:405 
\bibitem{Mic3}  Alexopoulos, T. et. al: A spark-resistant bulk-Micromegas chamber for high-rate applications, Nucl.Instrum.Meth., A640, 2011, PP:110-118
\bibitem{Mic4}  Alexopoulos, T. et. al: Development of large size Micromegas detector for the upgrade of the ATLAS muon system, Nucl.Instrum.Meth., A617, 2010, PP: 161-165
 
\end{thebibliography}
\end{document}